\documentclass[]{spie}  
\usepackage[]{graphicx,natbib}

\title{PSF reconstruction for NAOS-CONICA} 

\author{Yann Cl\'enet\supit{a}, Markus Kasper\supit{b}, Eric Gendron\supit{a}, Thierry Fusco\supit{c}, G\'erard Rousset\supit{a}, \\Damien Gratadour\supit{d}, Christopher Lidman\supit{e}, Olivier Marco\supit{e}, Nancy Ageorges\supit{e}, Sebastian Egner\supit{f}
\skiplinehalf
\supit{a}LESIA, Observatoire de Paris, 5 place Jules Janssen, 92195 Meudon cedex, France; \\
\supit{b}ESO, Karl-Schwarzschild-Strasse 2, D-85748 Garching bei M\"unchen, Germany; \\
\supit{c}ONERA, BP52, 29 avenue de la Division Leclerc, 92320 Ch\^{a}tillon Cedex, France;\\ 
\supit{d}Gemini Observatory, Gemini Observatory, Hilo, HI 96720, USA;\\
\supit{e}ESO, Alonso de C\'ordova, Casilla 3107, Vitacura, Casilla 19001, Santiago 19, Chile;\\
\supit{f}MPIA, K\"{o}nigstuhl 17, D-69117 Heidelberg, Germany;
}

\authorinfo{Further author information: send correspondence to Yann Cl\'enet (E-mail: yann.clenet@obspm.fr, Telephone: +33 1 45 07 75 48)}

\begin{document} 
  \maketitle 

\begin{abstract}
Adaptive optics (AO) allows one to derive the point spread function (PSF) simultaneously to the science image, which is a major advantage in post-processing tasks such as astrometry/photometry or deconvolution. Based on the algorithm of \citet{veran97}, PSF reconstruction has been developed for four different AO systems so far: PUEO, ALFA, Lick-AO and Altair. A similar effort is undertaken for NAOS/VLT in a collaboration between the group PHASE (Onera and Observatoire de Paris/LESIA) and ESO. In this paper, we first introduce two new algorithms that prevent the use of the so-called "$U_{ij}$ functions" to: (1) avoid the storage of a large amount of data (for both new algorithms), (2) shorten the PSF reconstruction computation time (for one of the two) and (3) provide an estimation of the PSF variability (for the other one). We then identify and explain issues in the exploitation of real-time Shack-Hartmann (SH) data for PSF reconstruction, emphasising the large impact of thresholding in the accuracy of the phase residual estimation. Finally, we present the data provided by the NAOS real-time computer (RTC) to reconstruct PSF ({\em (1)} the data presently available, {\em (2)} two NAOS software modifications that would provide new data to increase the accuracy of the PSF reconstruction and {\em (3)} the tests of these modifications) and the PSF reconstruction algorithms we are developing for NAOS on that basis.
\end{abstract}



\section{INTRODUCTION}
\label{sect:intro}  

The knowledge of the PSF of the instrument is of prime importance to accurately calibrate observations. Numerous image restoration techniques, such as deconvolution algorithms, make use of the PSF, or its Fourier transform, the optical transfert function (OTF), to fully restore the image quality affected by the atmopheric turbulence. Astrometric and photometric algorithms, e.g. DAOPHOT \citep{stetson87} or Starfinder \citep{diolaiti00}, sometimes coupled to deconvolution algorithms, also need an estimation of the PSF.

\citet{veran97} have demonstrated that AO allows one to derive the PSF from wavefront-related data delivered by the AO RTC. They have been the first to develop such a PSF reconstruction algorithm and to implement it on an AO system: PUEO, the CFHT curvature AO system \citep{arsenault94}. Since then, the algorithm is routinely used to provide astronomers with the (on-axis) PSF associated to their observations.

Based on this algorithm, PSF reconstruction has been developed for three other AO systems, equipped this time with SH wavefront sensors (WFS): (1) ALFA \citep{kasper00}, the SH AO system of the Calar Alto 3.5m telescope, by \citet{weiss03}; (2) Altair \citep{herriot2000}, the 4-quadrant SH AO system of the Gemini North telescope, by \citet{jolissaint04}; (3) the SH AO system of the UCO/Lick Observatory's 3 m Shane Telescope \citep{bauman02}, by \citet{fitz04}. These algorithms have been tested during few runs of observations and have lead to good results. 

With the goal to reconstruct PSF with NAOS, the SH AO system  of NACO  \citep{rousset00,lenzen98} at VLT, a similar effort has been undertaken in a collaboration between the group PHASE (Onera and Observatoire de Paris/LESIA) and ESO: PSF reconstruction had indeed been considered since the design phase of NAOS. In a first step, we have contacted MPIA (S. Egner and S. Hippler) since they had been the first to develop a PSF reconstruction algorithm for a SH AO system. Though, the large differences between the two systems (modes of the systems, available wavefront-related measurements) have lead us to elaborate a NAOS-dedicated algorithm, the general architecture being always based on the algorithm of \citet{veran97}.

In the first section, we introduce new algorithms for PSF reconstruction. In the second section, we identify and explain issues in the exploitation of real-time SH data for PSF reconstruction. In the third section, we describe the data provided by the NAOS RTC to reconstruct PSF and the contemplated software modifications that would provide new data to increase the accuracy of the PSF reconstruction. We also detail the algorithms associated with these data.

\section{New algorithms for PSF reconstruction}
\subsection{Reminder on the \citet{veran97} algorithm}
\label{sect:veran}
\subsubsection{The long-exposure AO-corrected PSF expression}
Following the PSF reconstruction algorithm developed by \citet{veran97}, assuming that the phase is quasi-stationary over the pupil, the AO-corrected monochromatic long-exposure OTF is decomposed as follows:
\begin{equation}
\Big\langle OTF\big(\vec{\rho}/\lambda\big)\Big\rangle =  \Big\langle OTF_{\phi_\epsilon}\big(\vec{\rho}/\lambda\big)\Big\rangle\times OTF_{\mathrm{tel}}\big(\vec{\rho}/\lambda\big)
\label{eq1}
\end{equation}
The phase $\phi_\epsilon$ can be split into two parts: $\phi_{\epsilon_\|}$, which belongs to the vector space spanned by the mirror modes, and $\phi_{\epsilon_\perp}$, which is orthogonal to the former space.
\begin{equation}
 \Big\langle OTF\big(\vec{\rho}/\lambda\big)\Big\rangle=\Big\langle OTF_{\phi_{\epsilon_\|}}\big(\vec{\rho}/\lambda\big)\Big\rangle\times\Big\langle OTF_{\phi_{\epsilon_\perp}}\big(\vec{\rho}/\lambda\big)\Big\rangle\times OTF_{\mathrm{tel}}\big(\vec{\rho}/\lambda\big)
\label{eq2}
\end{equation}
This expression can be written
\begin{equation}
 \Big\langle OTF\big(\vec{\rho}/\lambda\big)\Big\rangle = \exp\big(-\frac{1}{2} \bar{D}_{\phi_{\epsilon_\|}}(\vec{\rho})\big)\times\exp\big(-\frac{1}{2} \bar{D}_{\phi_{\epsilon_\perp}}(\vec{\rho})\big)\times OTF_{\mathrm{tel}}\big(\vec{\rho}/\lambda\big)
\label{eq3}
\end{equation}
\noindent where: \begin{itemize}
\item $\Big\langle OTF_{\phi_\epsilon}\big(\vec{\rho}/\lambda\big)\Big\rangle$ is the mean attenuation of the OTF due to the partial correction of AO,
\item $OTF_{\mathrm{tel}}\big(\vec{\rho}/\lambda\big)$ is the perfect telescope OTF,
\item $\Big\langle OTF_{\phi_{\epsilon_\|}}\big(\vec{\rho}/\lambda\big)\Big\rangle$ is the mean attenuation of the OTF due the mirror component of the phase, i.e. the "mean residual phase OTF",
\item $\Big\langle OTF_{\phi_{\epsilon_\perp}}\big(\vec{\rho}/\lambda\big)\Big\rangle$ is the mean attenuation of the OTF due the component of the phase belonging to the space perpendicular to the mirror space, i.e. the "mean perpendicular phase OTF", 
\item $\bar{D}_{\phi_{\epsilon_\|}}(\vec{\rho})$ is the mean structure function of the residual phase, i.e. the "mean residual phase structure function",
\item $\bar{D}_{\phi_{\epsilon_\perp}}(\vec{\rho})$ is the mean structure function of the perpendicular phase,
\item $\vec{\rho}$ is a pupil plane coordinate vector,
\item $\lambda$ is the wavelength of observation.
\end{itemize}
The corresponding  AO-corrected monochromatic long-exposure PSF is derived as the Fourier transform of the OTF.

From the expression of the residual phase structure function:
\begin{equation}
D_{\phi_{\epsilon_\|}}(\vec{x},\vec{\rho})=\Big\langle\big(\phi_{\epsilon_\|}(\vec{x})-\phi_{\epsilon_\|}(\vec{x}+\vec{\rho})\big)^2\Big\rangle
\label{eq4}
\end{equation}
\noindent and the decomposition of the phase on the basis of the mirror modes $M_i(\vec{x})$: 
\begin{equation}
\phi_{\epsilon_\|}(\vec{x},t)=\sum_{i=1}^N \epsilon_{\|i}(t)\,M_i(\vec{x})
\label{eq5}
\end{equation}
\noindent one obtains:
\begin{equation}
D_{\phi_{\epsilon_\|}}(\vec{x},\vec{\rho})=\sum_{i=1}^N\sum_{j=1}^N \langle\epsilon_{\|i}\epsilon_{\|j}\rangle\big(M_i(\vec{x})-M_i(\vec{x}+\vec{\rho})\big)\big(M_j(\vec{x})-M_j(\vec{x}+\vec{\rho})\big)
\label{eq6}
\end{equation}
\noindent where $\vec{x}$ a coordinate vector in the pupil plane.

The mean residual phase structure function $\bar{D}_{\phi_{\epsilon_\|}}(\vec{\rho})$ is the mean of $D_{\phi_{\epsilon_\|}}(\vec{x},\vec{\rho})$ over $\vec{x}$:
\begin{equation}
\bar{D}_{\phi_{\epsilon_\|}}(\vec{\rho})=\frac{\displaystyle{\int D_{\phi_{\epsilon_\|}}(\vec{x},\vec{\rho})P(\vec{x}) P(\vec{x}+\vec{\rho})\mathrm{d}\vec{x}}}{\displaystyle{\int P(\vec{x})\, P(\vec{x}+\vec{\rho})\mathrm{d}\vec{x}}}
\label{eq7}
\end{equation}
\noindent where $P(\vec{x})$ is the pupil function.

\subsubsection{The $U_{ij}(\vec{\rho})$ functions}

Defining the function $U_{ij}(\vec{\rho})$ as:
\begin{equation}
\frac{\displaystyle{\int \!\big(M_i(\vec{x})-M_i(\vec{x}+\vec{\rho})\big)\big(M_j(\vec{x})-M_j(\vec{x}+\vec{\rho})\big)P(\vec{x}) P(\vec{x}+\vec{\rho})\mathrm{d}\vec{x}}}{\displaystyle{\int P(\vec{x})\, P(\vec{x}+\vec{\rho})\mathrm{d}\vec{x}}}
\label{eq9}
\end{equation}
\noindent Eq.~\ref{eq7}, can be rewritten:
\begin{equation}
\bar{D}_{\phi_{\epsilon_\|}}(\vec{\rho})=\sum_{i=1}^N\sum_{j=1}^N \langle\epsilon_{\|i}\epsilon_{\|j}\rangle\,U_{ij}(\vec{\rho})
\label{eq8}
\end{equation}
This is a key equation for the experimental PSF reconstruction. The covariance matrix $\langle\epsilon_\|{\epsilon_\|}^t\rangle$ has to be measured experimentally on the AO system itself, by averaging the cross-products of wavefront measurements obtained during the time of the image exposure.  

In the current PSF reconstruction algorithms, derived from \citet{veran97}, the matrix $\langle\epsilon_\|{\epsilon_\|}^t\rangle$ is the basic entry point from which one can deduce successively the phase structure function, the OTF, and then the PSF. Additionally, one has to compute, store once for all, and also read during the reconstruction process the $U_{ij}(\vec{\rho})$ functions.

In Eqs.~\ref{eq8}, the $i$ and $j$ indices play a symmetric role, so that there are actually $N\times(N+1)/2$ "useful" $U_{ij}(\vec{\rho})$ functions. As an example, in the case of  NAOS, the 159 compensated modes lead to 12720 "useful" $U_{ij}(\vec{\rho})$ functions. Today, the large number of $U_{ij}(\vec{\rho})$ hence represents, depending on the array size and data type, several gigabytes of data to compute, store and read. This leads to a heavy PSF reconstruction process, which will turn out to be impossible to handle in the future since next AO systems are expected to have a largely increased number of modes: about 1370 actuators for the VLT-Planet Finder AO system \citep{fusco05},  several tens of thousands for extremely large telescopes.  We propose in the following a way to achieve this computation, starting from the same covariance matrix $\langle\epsilon_\|{\epsilon_\|}^t\rangle$, without using the $U_{ij}(\vec{\rho})$.

\subsection{Theory of the proposed algorithms}
\label{sect:new_algo}
\subsubsection{The $V_{ii}$ algorithm}

Let's consider the vector $\epsilon_\|$\footnote{to lighten the expressions, $\epsilon_\|(t)$ will be abbreviated into $\epsilon_\|$} made of the $\{\epsilon_{\|i}\}_{i=1...N}$ coefficients, i.e. the vector representing $\phi_{\epsilon_\|}(\vec{x},t)$ in the basis of the mirror modes $M_i(\vec{x})$. The eigen decomposition of the residual phase covariance matrix is:
\begin{equation}
\Lambda=B^t\langle\epsilon_\|{\epsilon_\|}^t\rangle B
\label{eq10}
\end{equation}
\noindent where $\Lambda$ is a diagonal matrix that contains the $\{\lambda_i\}_{i=1...N}$ eigenvalues and B is the matrix of eigenvectors: $B^tB=BB^t=Id$. Equation~\ref{eq10} can be written:
\begin{equation}
\Lambda=\big\langle(B^t\epsilon_\|)(B^t\epsilon_\|)^t\big\rangle
\label{eq11}
\end{equation}
The vector $\eta$ equal to  $B^t\epsilon_\|$ represents $\phi_{\epsilon_\|}(\vec{x},t)$ in the basis that diagonalizes the residual phase covariance matrix. Its coefficients are noted $\{\eta_i\}_{i=1...N}$. From Eq.~\ref{eq11}, the covariance matrix $\langle\eta \eta^t\rangle$ is diagonal, i.e. in this new basis, the residual phase covariance matrix is diagonal and the mean residual phase structure function reduces to:
\begin{equation}
\bar{D}_{\phi_{\epsilon_\|}}(\vec{\rho})=\sum_{i=1}^N \langle\eta_i\eta_i\rangle\,V_{ii}(\vec{\rho})=\sum_{i=1}^N \lambda_i\,V_{ii}(\vec{\rho})
\label{eq12}
\end{equation}
\noindent where the $V_{ij}(\vec{\rho})$ functions are equivalent in the new basis to the $U_{ij}(\vec{\rho})$ functions (Eq.~\ref{eq9}).

After this change of basis, the computation of the residual phase OTF only requires the computation of a number $N$ of functions $V_{ii}(\vec{\rho})$. Though, these $V_{ii}(\vec{\rho})$ functions have to be computed on the fly for each estimation of the mean residual phase structure function.

\subsubsection{The "instantaneous PSF" algorithm}

The solution we propose here is similar to the algorithm presented by \citet{roddier90} to simulate atmospherically distorted wavefronts, from the covariance matrix of the coefficients of their expansion in Zernike modes. We extend this algorithm to any modal basis and covariance matrix, and use it to reconstruct AO-corrected PSFs.

Let's consider again the eigen decomposition of the residual phase covariance matrix:
\begin{equation}
\langle\epsilon_\|{\epsilon_\|}^t\rangle=B\Lambda B^t
\label{eq13}
\end{equation}
If one generates a vector $\eta$ whose coefficients are independent Gaussian random variables with zero mean and variance equal to the eigenvalue $\lambda_i$, i.e. $\langle \eta \eta^t\rangle=\Lambda$,
then the vector $\beta=B\eta$ is a set of correlated random variables whose covariance matrix is  $\langle\epsilon_\|{\epsilon_\|}^t\rangle$: 
\begin{equation}
\langle \beta\beta^t\rangle=\langle B\eta\eta^tB^t\rangle= B\Lambda B^t=\langle\epsilon_\|{\epsilon_\|}^t\rangle
\label{eq14}
\end{equation}
The phase represented by the vector $ \beta$ is:
\begin{equation}
\phi(\vec{x},t)=\sum_{i=1}^N \beta_i(t)\,M_i(\vec{x})
\end{equation}
\noindent and the "instantaneous" PSF corresponding to that phase is:
\begin{equation}
PSF_\|(\vec{x},t)=\Big\|\mathcal{FFT}\big(\exp(i\phi(\vec{x},t)\big)\Big\|^2
\end{equation}
Then, by generating random $\eta$ vectors such that $\langle \eta \eta^t\rangle=\Lambda$, we build instantaneous PSFs that, in average, converge to the long-exposure PSF of the mirror space. Note that the latter is not the "full PSF" as would be observed at the telescope since it does not include the uncorrected part of the phase (cf. Eq.~\ref{eq2}).

\subsection{Choice between the different algorithms}

The $U_{ij}$ and $V_{ii}$ algorithms mathematically produce exactly the same OTFs. Though, the $V_{ii}$ algorithm requires the computation of $N$ functions (where $N$ is the number of modes) whereas the  $U_{ij}$ algorithm requires the computation of $N\times(N+1)/2$ functions. Even if these $N$ $V_{ii}$ functions have to be computed on the fly, this computation is faster than reading  $N\times(N+1)/2$ stored $U_{ij}$ functions and it prevents from the storage of a large amount of data, so that the $V_{ii}$ algorithm is always to be preferred to the $U_{ij}$ one.

As noticed by \citet{conan94}, averaging short-exposure OTFs, as we do in the "instantaneous PSF" algorithm, is a process that converges very slowly, especially at large $D/r_0$ or low correction level. In addition, it does not lead to the infinitely long exposure OTF since a given number of short exposures are averaged. Besides, \citet{conan94} has shown that in the poor correction case, the error in computing the long-exposure OTF of such an algorithm is larger than for the $U_{ij}$ algorithm, and consequently the $V_{ii}$ algorithm as well.

Though, we emphasize that, in addition to the OTF computation itself, the "instantaneous PSF" algorithm can provide an estimation about the variability of the OTF,  that can help a lot in some deconvolution algorithms. The estimation of the infinitely long exposure OTF that results from the convergence of our "instantaneous PSF" algorithm and that corresponds to a given covariance matrix $\langle\epsilon_\|{\epsilon_\|}^t\rangle$ is unique: let us call it $OTF_{\infty}(\vec{\rho}/\lambda)$. The dispersion in the random generation of OTFs can be computed as 
\begin{equation}
\sigma^2(\vec{\rho}/\lambda) = \Big\langle \big\|OTF_{\infty}(\vec{\rho}/\lambda) - OTF_{i}(\vec{\rho}/\lambda) \big\|^2 \Big\rangle_{i}
\end{equation} 
where $OTF_i$ is the $i^{th}$ draw of a randomly-generated OTF.
If we call $OTF_\mathrm{{obs}}(\vec{\rho}/\lambda)$ the OTF actually observed on the instrument during the given, non infinite, observing time $T_\mathrm{{int}}$, when the given covariance matrix $\langle\epsilon_\|{\epsilon_\|}^t\rangle$ was measured, we can evaluate how far our estimation $OTF_{\infty}$ is from $OTF_\mathrm{{obs}}$ by writing:
\begin{equation}
\Big\langle \big\|OTF_{\infty}(\vec{\rho}/\lambda) - OTF_\mathrm{{obs}}(\vec{\rho}/\lambda) \big\|^2 \Big\rangle = 
\frac{\sigma^2(\vec{\rho}/\lambda)}{n}
\end{equation}
where $n$ is the equivalent number of independent realisations of PSFs, whose sum has resulted in the final PSF observed by the instrument during the given, non infinite, integration time $T_\mathrm{{int}}$. An estimation of $n$ could be obtained for example from a full simulation of the AO system under the same atmospheric conditions as during the observation. It can also be reasonnable to consider that the impact of the correction by the AO system is to shorten the image correlation time compared to the image correlation time $\tau_0(\lambda)$ of the atmospheric seeing \citep{rigaut91}. We can then find a lower bound given by $n> T_\mathrm{{int}}/\tau_0(\lambda)$.

\section{Issues in the exploitation of real-time Shack-Hartmann data for PSF reconstruction}
\subsection{Description of the problem}
\label{section:yao_onera_simu}
\subsubsection{Using the Yao simulation software}
\begin{figure}[t]
\centering
   \begin{tabular}{cc}
     \includegraphics[width=6cm]{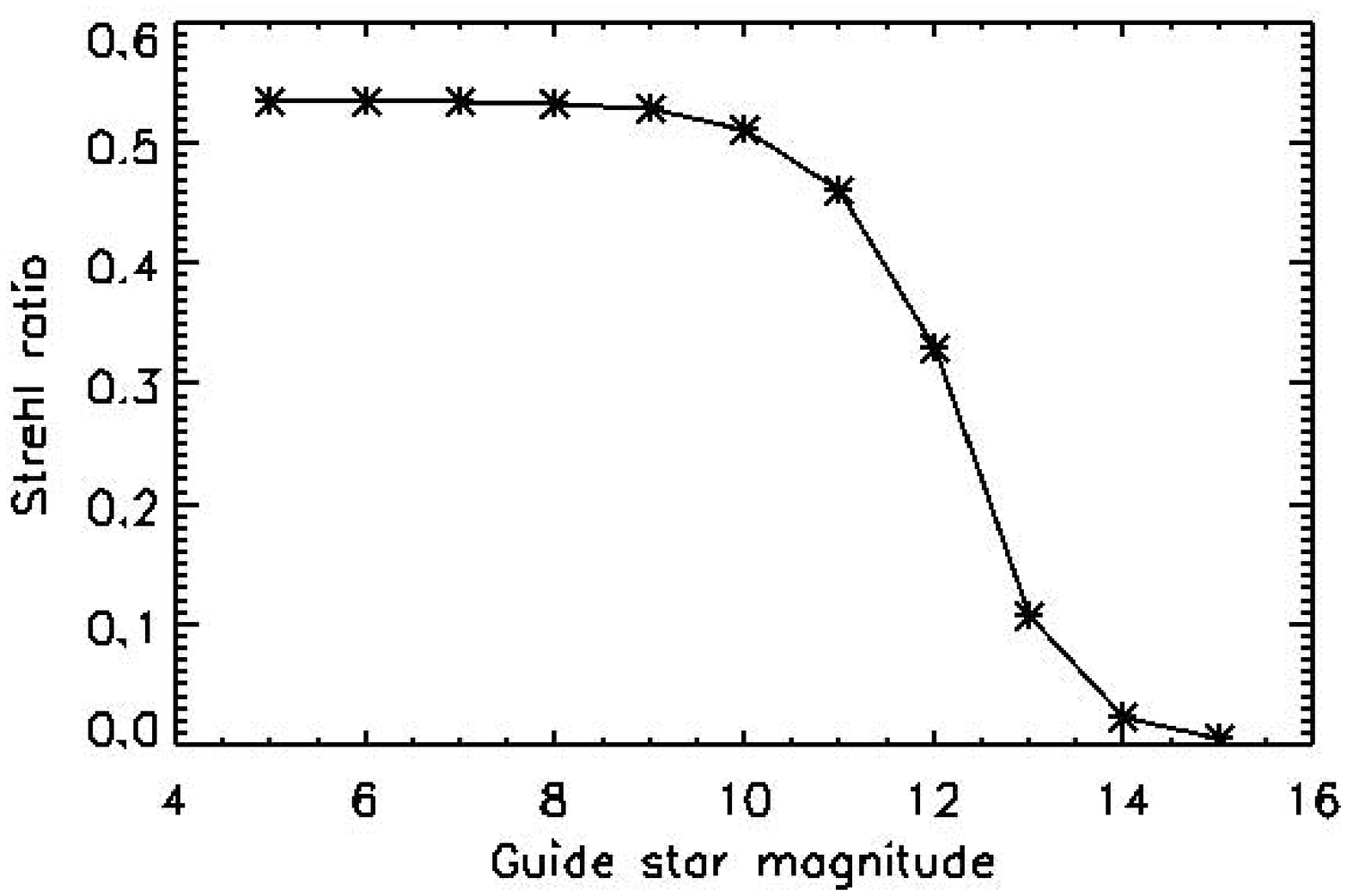}&
    \includegraphics[width=6cm]{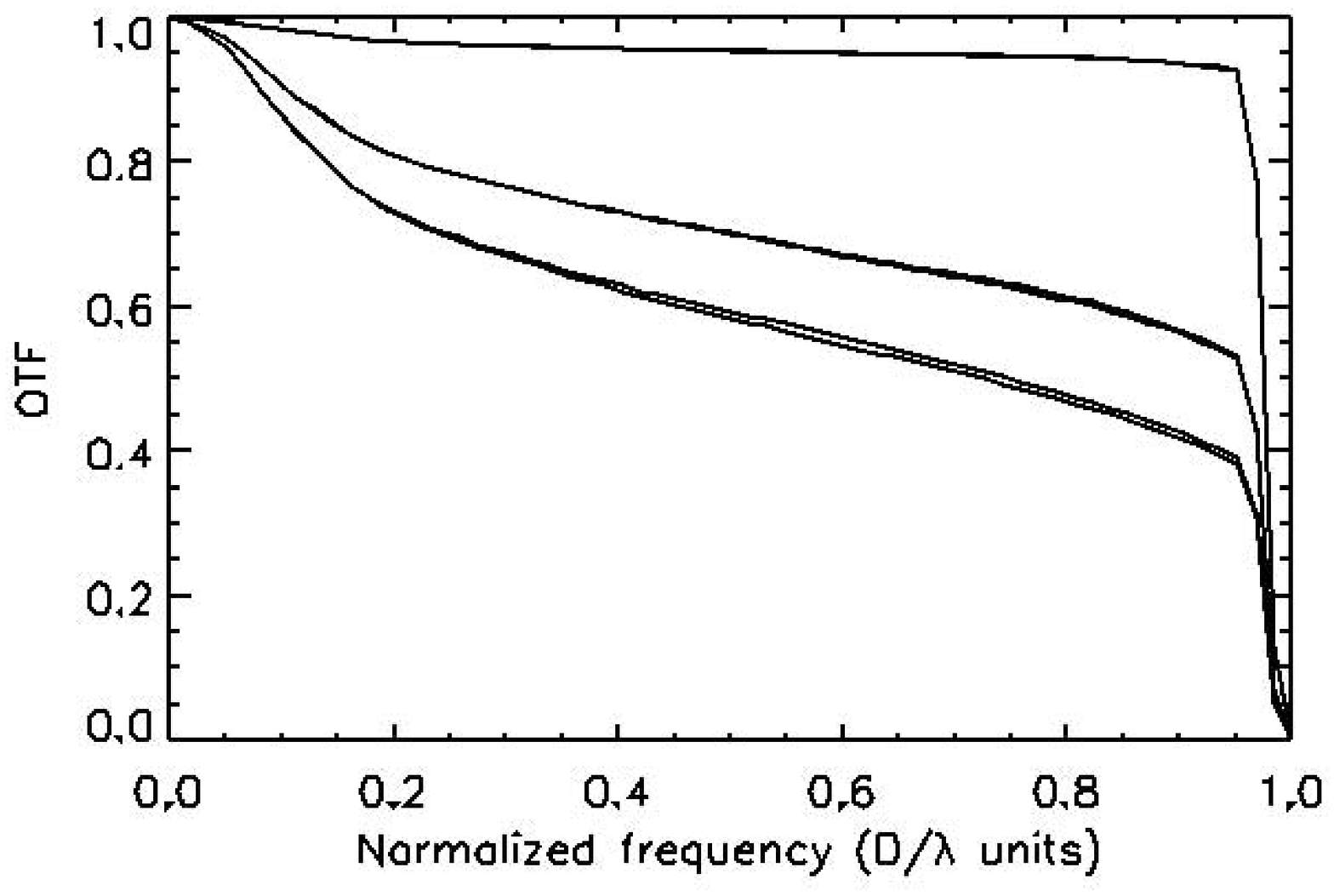}   
   \end{tabular}
      \caption{Left: Strehl ratio at 1.65 $\mu m$ vs. guide star V magnitude, from Yao. Yao computes the Strehl ratios from the PSFs resulting from the simulation (as would be observed on a camera). Right: Circular mean of the "atmospheric" OTFs for guide star magnitudes of 5, 12 and 14 (from top to bottom), computed from the WFS data. At this scale, the profiles obtained with the different algorithms ($U_{ij}$, $V_{ii}$ and "instantaneous PSF") are almost superimposed.}
         \label{fig:yao}
\end{figure}
In order to test the algorithms introduced in Sect.~\ref{sect:new_algo}, we have used the end-to-end AO simulation software Yao (version 3.6), written by  \citet{rigaut05}. We have tested our algorithms in the simple case of  a 6$\times$6 subpupil SH by using the default configuration file "sh6m2-bench.par" delivered in the Yao installation package. The correction was performed with a 45 actuator piezostack deformable mirror (DM) plus a tip-tilt mirror. The correction and observation wavelengths of the simulation were 0.65 and 1.65 $\mu$m respectively. $D/r_0$ was equal to 42.44 at 0.5 $\mu$m.

Since we aimed at testing our algorithms with different conditions of correction, we have run the simulation with a guide star magnitude ranging from 5 to 14 so that the resulting Strehl ratio was ranging from 0.54 down to 0.01 (Fig.~\ref{fig:yao} left). We have stored all the values  $\epsilon_\|(t)$ obtained from the simulation, computed the  covariance matrix $\langle\epsilon_\|{\epsilon_\|}^t\rangle$ and ran the $U_{ij}$, $V_{ii}$  and "instantaneous PSF" algorithms to derive the corresponding "atmospheric" OTFs (i.e. not multiplied by the telescope OTF; Fig.~\ref{fig:yao} right).

The Strehl ratios that one would expect from the OTF profiles (Fig.~\ref{fig:yao} right), i.e. computed from the WFS data, do not correspond to the Strehl ratios computed from the PSFs resulting from the simulation (Fig.~\ref{fig:yao} left), even if one takes into account that these OTFs only correspond to the residual phase, without correction for the noise, the aliasing and the perpendicular parts of the phase. This is particularly evident for the 14$^{th}$ guide star magnitude case, for which one would expect much smaller OTF values for a large range of normalized frequencies. In order to understand this discrepancy, we have reproduced the test with another AO simulation software.

\subsubsection{Using the ONERA AO simulation software}
To trace the origin of the problem, we have tested the algorithm with an end-to-end Monte Carlo-based AO simulation software developed at ONERA \citep{conan04}. We have tested our algorithms in the simple case of  a NAOS-like AO system
\citep{rousset00}: a 14$\times$14 subpupil SH with 8$\times$8 pixels per sub-aperture, a read-out noise of 3 e$^-$ per pixel and a sampling frequency of  500 Hz. The correction was performed with a 185 actuator piezostack DM plus a tip-tilt mirror. The correction and observation wavelengths of the simulation were 0.65 and 2.2 $\mu$m respectively. The seeing was 0.85" at 0.5~$\mu$m. 

\begin{figure}[t]
   \centering
   \begin{tabular}{cc}
    \includegraphics[width=6cm]{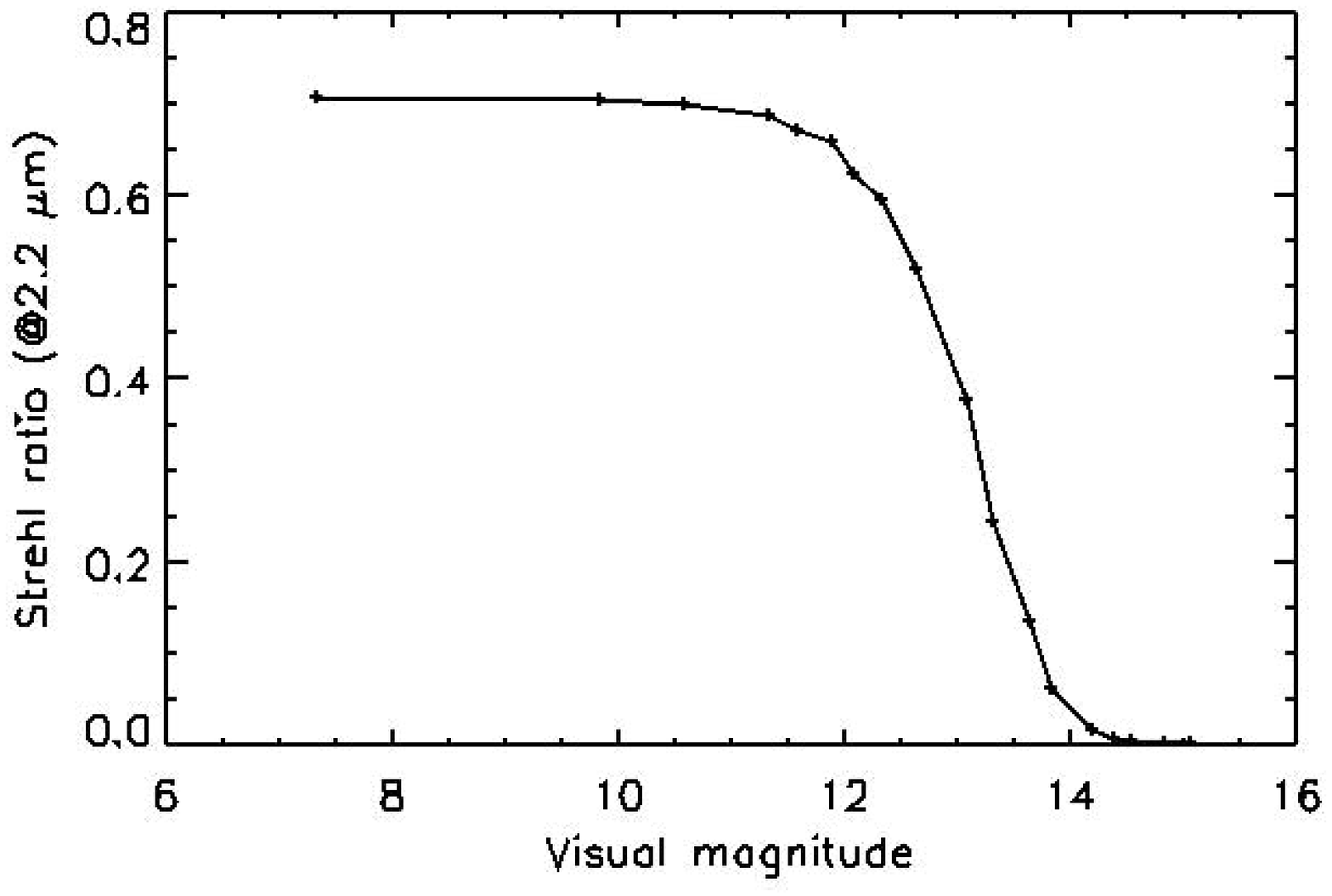}&
    \includegraphics[width=6cm]{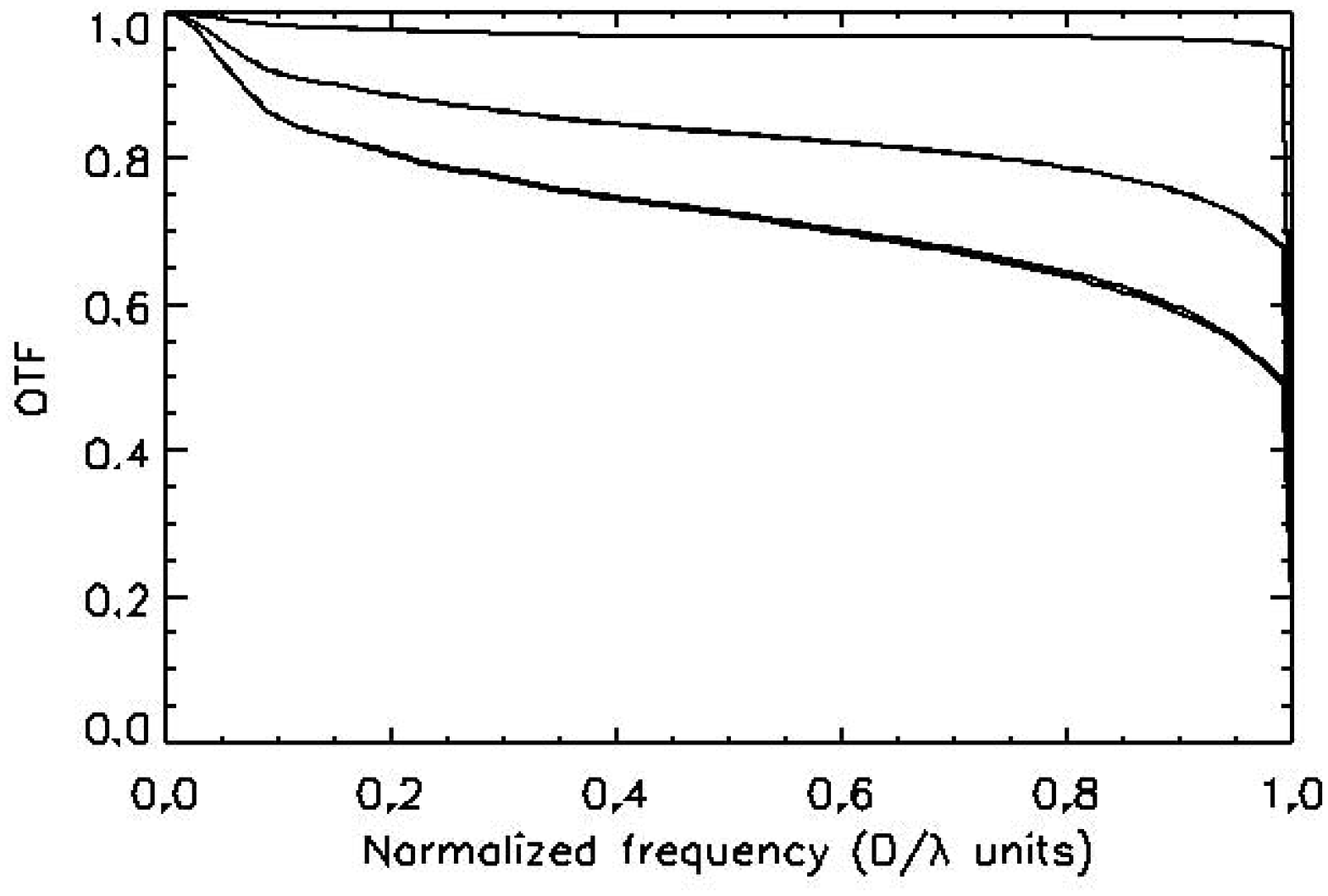}
    \end{tabular}
      \caption{Left: Strehl ratio at 2.2 $\mu$m vs. guide star visual magnitude for a NAOS-like AO system. As for Yao, the Strehl ratios are computed from the PSFs obtained from the simulation (as would be seen in the "imaging path"). Right: Circular mean of the "atmospheric" residual OTFs for guide star magnitudes of 7.3, 12.3 and 13.6 (from top to bottom), computed from the WFS data. At this scale, the profiles obtained with the different algorithms ($U_{ij}$, $V_{ii}$ and "instantaneous PSF") are almost superimposed.}
         \label{fig:simu2}
   \end{figure}

Similarly to what we have done with Yao, we have run the simulation with a guide star magnitude ranging from 7 to 15, leading to a resulting Strehl ratio from $\sim$70\%
down to $\sim$0.2\% (Fig.~\ref{fig:simu2} left), and for a given guide star magnitude, we have stored the values  $\epsilon_\|(t)$ obtained from the simulation, computed the  covariance matrix $\langle\epsilon_\|{\epsilon_\|}^t\rangle$ and run the $U_{ij}$, $V_{ii}$  and "instantaneous PSF" algorithms to derive the corresponding "atmospheric" OTFs (Fig.~\ref{fig:simu2} right).

The behaviour we had encountered with Yao has still been observed in this other simulation: there exists a discrepancy between the Strehl ratios computed from the "imaging path"-related data (the PSFs, left graph of Fig.~\ref{fig:simu2}) and the Strehl ratios expected from the WFS data (the OTFs, right graph of Fig.~\ref{fig:simu2}).

\subsection{Origin of the problem}
We have identified unambiguously that the unexpected high OTF values are related to the correlation between the error and the turbulence signal.

In order to analyse the effect in more detail, we have simulated the behaviour of the SH measurement in one subaperture in particular. We simulate the incoming wavefront, compute the associated image with a high resolution and derive the position of its centre of gravity (hereafter the "true centroid position"). Then, we simulate the technical characteristics of the sensor, namely: the averaging of the image over the pixel area, the limited number of pixels and field of view, the detector noise, the signal photon noise. We apply a threshold on the image and compute the centre of gravity (hereafter the "measured centroid position"). After comparing the measured centroid position ($s_m$, for measured slope) with the true one ($s_0$), it appears clearly that: 
\begin{itemize} 
\item the error $s_m-s_0$ is highly correlated to the signal value (i.e. to the true centroid position), which is contradictory with the assumptions usually made in the literature;
\item the correlation is, among others, a function of the threshold level.
\end{itemize}
Therefore, instead of the usual equation $s_m = s_0 + n$, the measured signal in a subaperture should rather be written under the form:
\begin{equation}
\label{G}
s_m = G.s_0 + n
\end{equation}
\noindent where $G$ is a coefficient and $n$ an additive noise, uncorrelated with the signal $s_0$. The value of $G$ can be found with a linear regression (least-squares fit by a function $ax+b$) of the measured data $s_m$ with respect to $s_0$. The fit residuals (i.e. $n$) are, de facto, uncorrelated with $s_0$, as a consequence of the least-squares approach. In addition, it should be noticed that $n$ may not have a zero mean, depending on the location of the reference slope in the subaperture.
The value of $G$ is related to the noisy pixel values that peak above the threshold and bias the estimation of the centroid location, because they weight in the denominator of the expression of the centre of gravity, given by (in the x-measurement case):
\begin{equation}
s_{m} = \frac{\displaystyle\sum_{i} x_i . ( S_i + N_i)}{\displaystyle\sum_{i} S_i + N_i}
 =  \left( \frac{\displaystyle\sum_{i}  S_i }{\displaystyle\sum_{i} S_i + N_i}\right)  . \left(\frac{\displaystyle\sum_{i} x_i .S_i}{\displaystyle\sum_{i} S_i}\right) + \frac{\displaystyle\sum_{i} x_i .  N_i}{\displaystyle\sum_{i} S_i + N_i}
\label{eq21}
\end{equation}

\noindent where $S_i$, $N_i$ and $x_i$ are the flux signal, the flux noise and the position of the pixel $i$ in the given supaperture. Equation~\ref{eq21} is comparable to Eq.~\ref{G}, where the mean value of $G$ can be expressed in terms of the total flux  $N_{phot}$ detected above the threshold, and in terms of the sum of the noise $N_{noise}$ {\em above the threshold level}
\begin{equation}
\label{pointG}
G \approx \frac{N_{phot}}{N_{phot}+N_{noise}}
\end{equation}

\begin{figure}[t]
   \centering
   \begin{tabular}{c c c c c}
   \includegraphics[width=4.5cm]{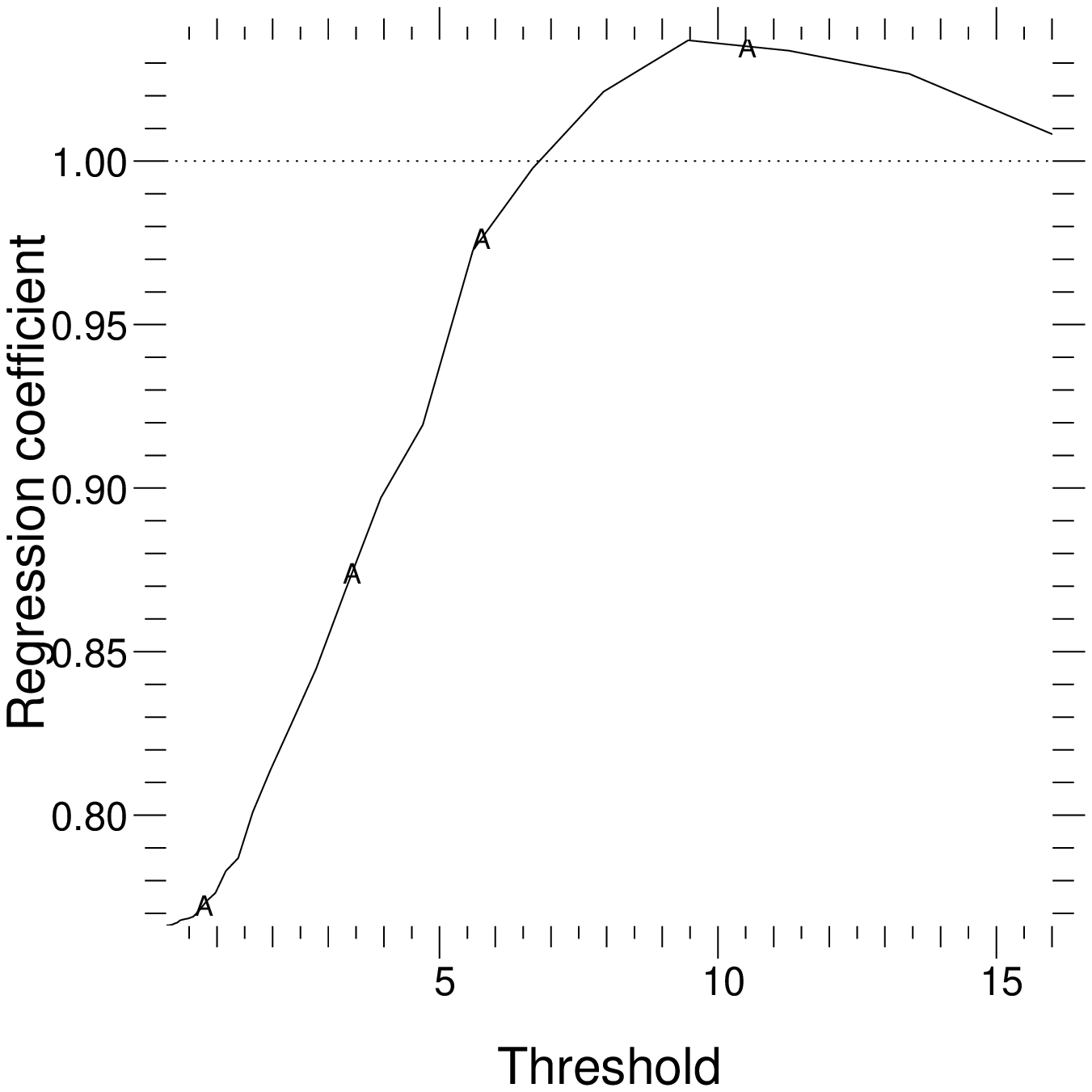}& &
   \includegraphics[width=4.5cm]{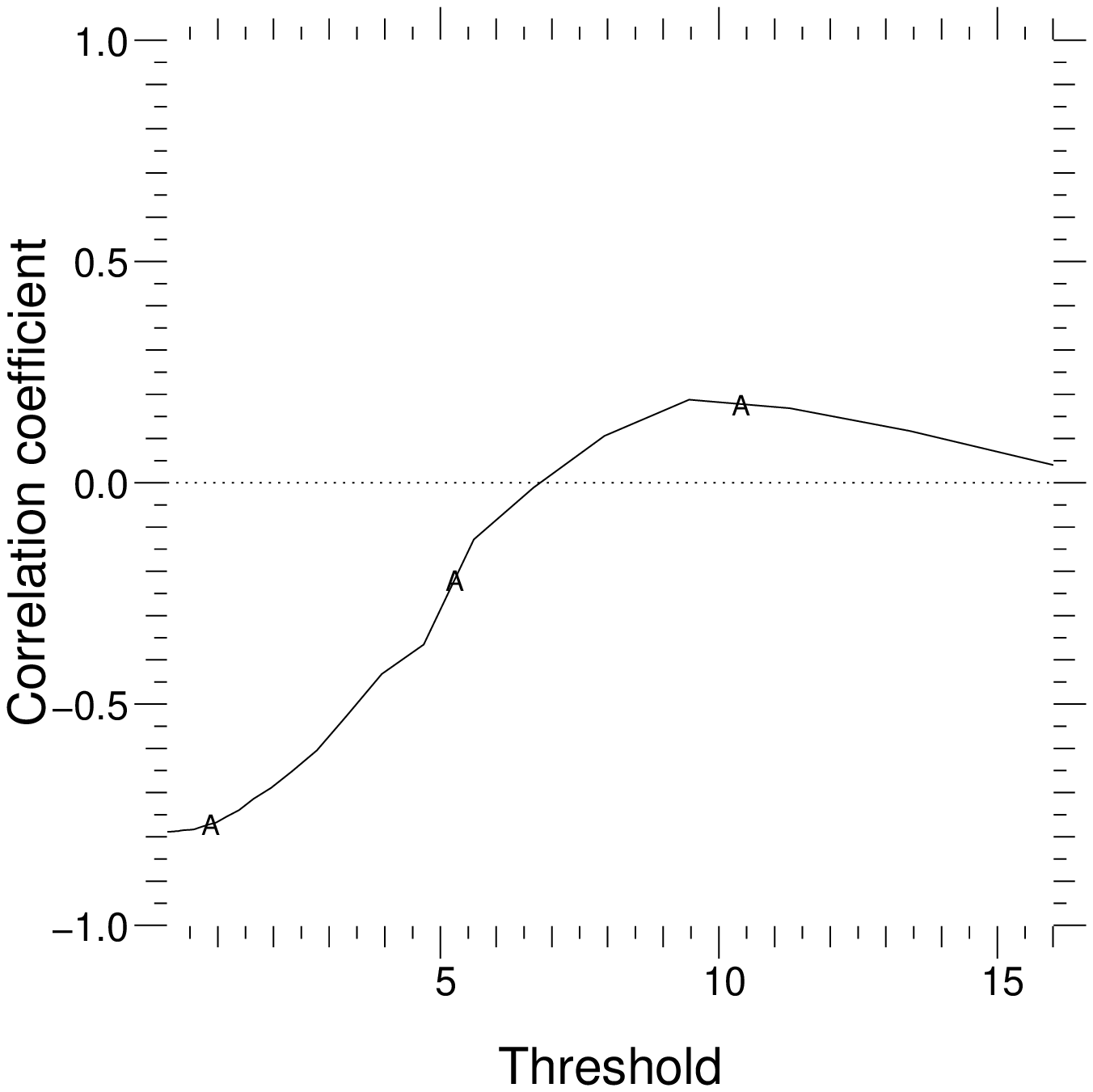}& & 
      \includegraphics[width=4.5cm]{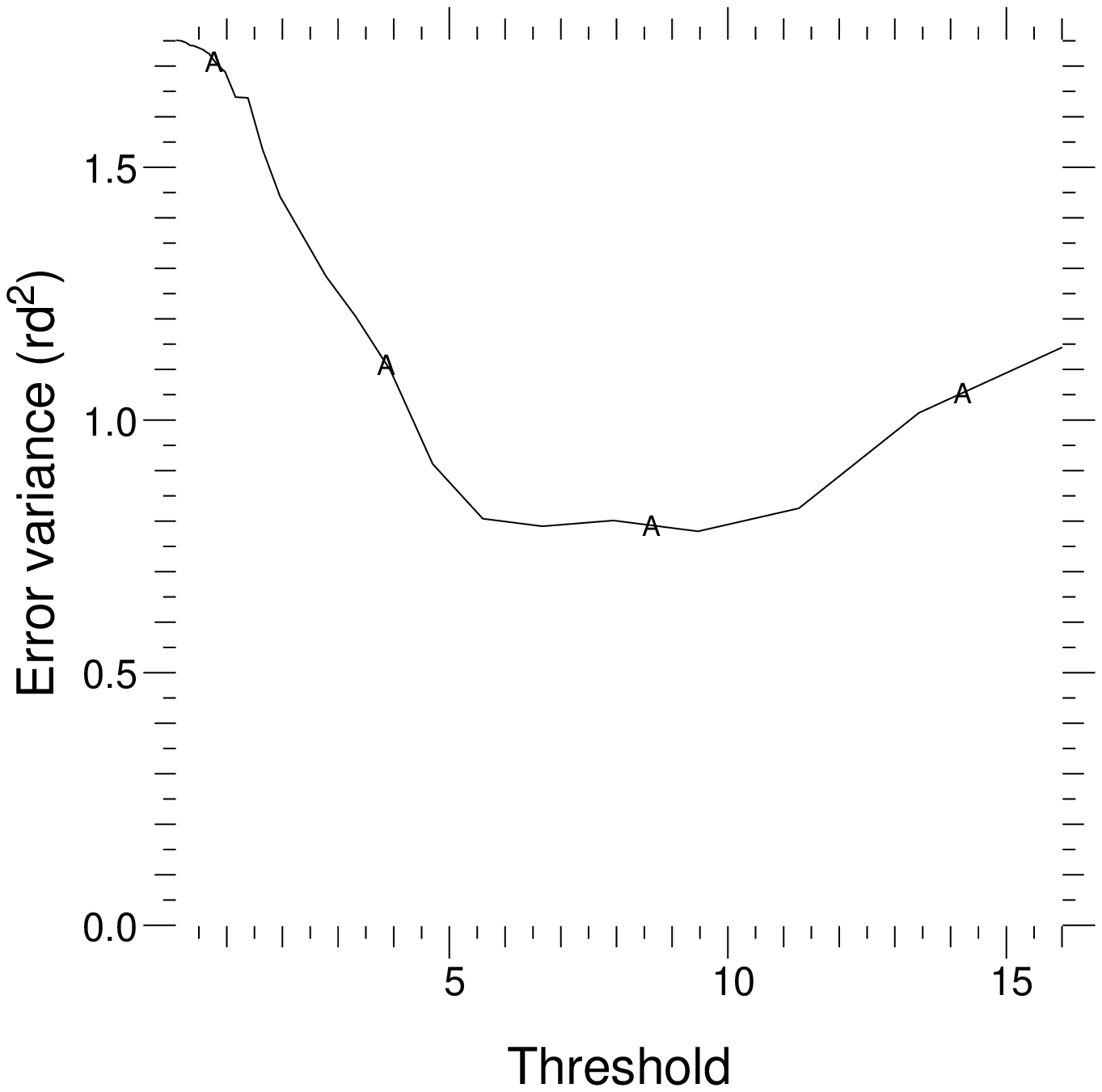}\\
      \end{tabular}
      \caption{Left : Coefficient $G$ vs. threshold from the linear regression of the measured data $s_m$ through $G.s_0 + constant$. Middle : Correlation coefficient between the signal $s_0$ and the error $s_m-s_0$ vs. threshold. Right : Variance of the error $\langle (s_m-s_0)^2 \rangle$ vs. threshold. Conditions for the 3 figures: $7\times 7$ subapertures, 8-m telescope, $r_0=13$ cm at 0.6 microns, detector read-out noise: 3 e$^-$, pixel scale: 0.4"/pixel, centroid computed on $6\times 6$ pixels, sampling frequency: 50 Hz, star magnitude: R=15, $\Delta \lambda=300$ nm, global throughput: 0.25.}
         \label{regress}
   \end{figure}

We have plotted in Fig.~\ref{regress} the value of $G$  versus the threshold level. While Eq.~\ref{pointG} can closely predict the value of $G$ for the low threshold values, it cannot predict the behavior for higher threshold levels, where the simulations show that, in some cases, we may even have $G>1$. The interesting point is that the error reaches a minimum when the correlation is zero, which coincides with a regression coefficient of $G=1$.

\begin{figure}[t]
   \centering
   \begin{tabular}{cc}
   \includegraphics[width=6cm]{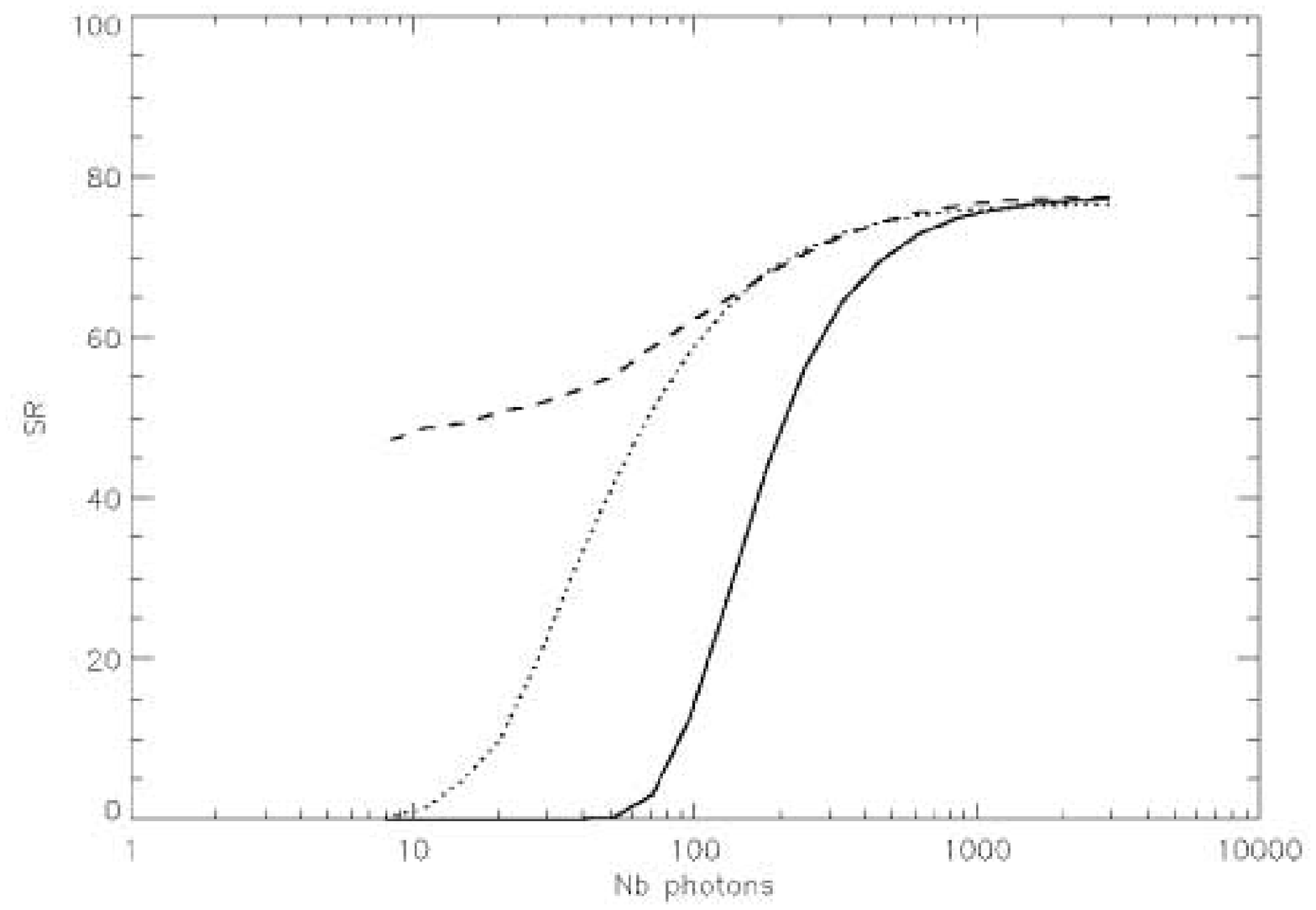}
   \includegraphics[width=6cm]{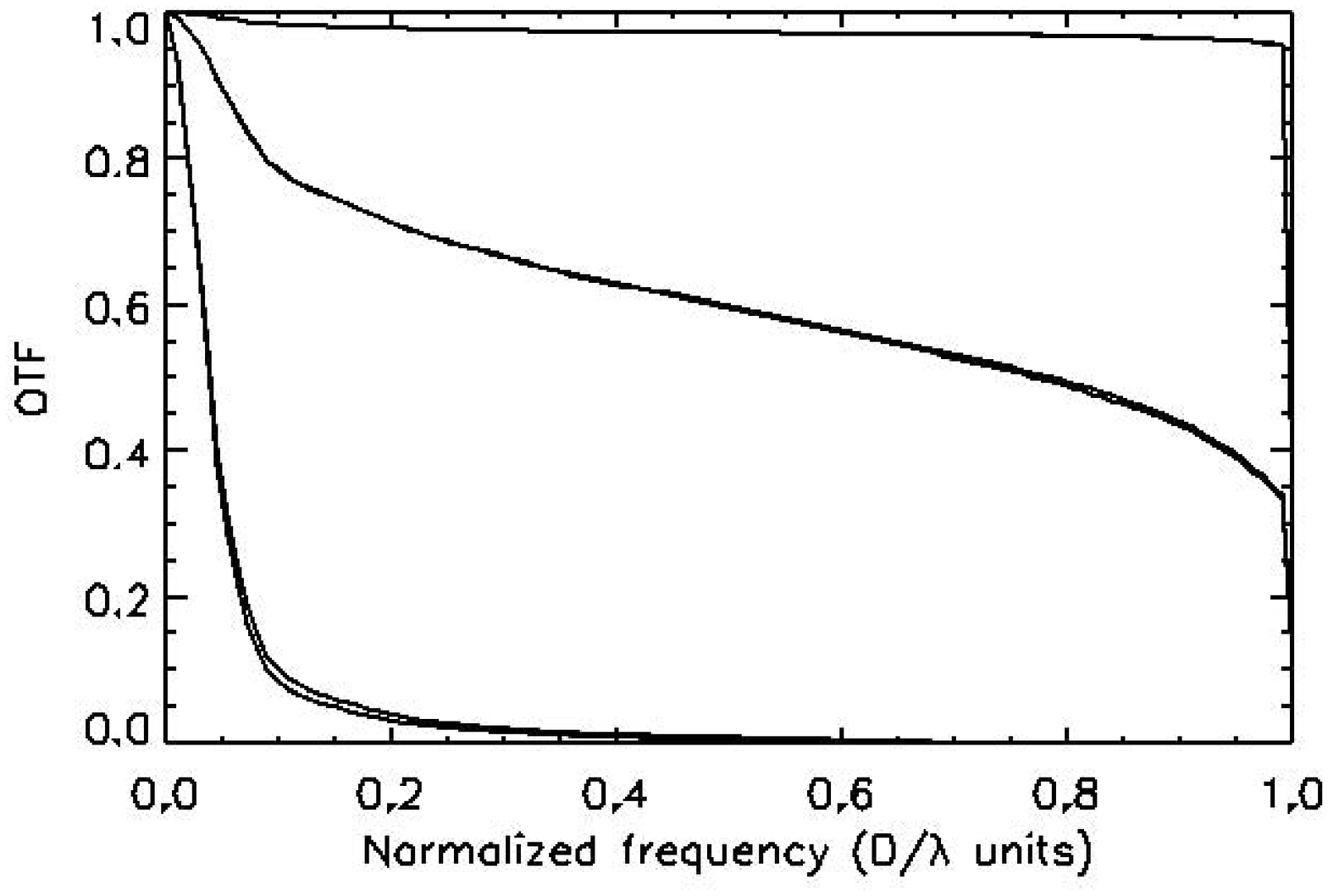}
   \end{tabular}
      \caption{Left: Strehl ratio vs. guide star flux (in number of photons). The slope is here computed as a simple phase difference at the subpupil edges and (1) we reproduce the threshold effect (upper dashed curve) and (2) we correct the slope measurements from the gain of the centroid measurement, but not from the noise (central dotted curve). These Strehl ratios are computed from WFS data for these curves and from "imaging path"-related data for the lower continuous one. Right: Same as Fig.~\ref{fig:simu2} right,  after correcting the slope measurements from the gain of the centroid measurement.}
         \label{fig:shgeo}
   \end{figure}

We have reproduced this particular "gain effect", encountered in our end-to-end simulations described in Sect.~\ref{section:yao_onera_simu}, using the ONERA AO simulation software, not in its end-to-end configuration but: \begin{itemize}
\item the slope is computed as a simple phase difference at the edge of the corresponding subpupil,
\item we then multiply by a gain factor $G=N_{phot}/\left(N_{phot}+N_{noise}\right)$, where $N_{noise}$ follows a truncated Gaussian statistics in order to reproduce the threshold effect.
\item the noise is then added to the slope measurement.
\end{itemize}
Indeed, the resulting upper dashed curve on the left graph of Fig.~\ref{fig:shgeo} reproduces the behaviour of our two end-to-end full-propagation simulations: the Strehl ratio reaches a lower limit at low flux instead of getting down to zero as observed from the "imaging path"-related data (lower continuous curve on  the left graph of Fig.~\ref{fig:shgeo}). However, if we correct the slope measurements from the gain of the centroid measurement, we find the expected behaviour (central dotted curve on the left graph of Fig.~\ref{fig:shgeo}, which is not superimposed on the lower continuous curve because the slope measurements have not been corrected for the noise).
   
As a conclusion, the threshold level is a key issue when reconstructing PSFs from real SH data. An improper threshold level may bias the data, leading to a wrong estimation of the noise and of the phase residuals. This issue may also impact on all the tasks that make use of the slope measurements, such as performance estimations (as done for NAOS), modal optimisation, etc. Surprisingly, its impact on the performance of the AO compensation itself is maybe less sensitive, since the gain $G$ acts exactly as the close-loop gain would do: it directly controls the system bandwidth, and only makes the system "slower" when $G<1$.

\section{NAOS data for PSF reconstruction and associated algorithms}
\subsection{Data presently available}
\label{sect:naosdata}
The delivery of wavefront-related data to estimate the PSF has been among the NAOS main specifications to maximise the scientific returns of the instrument. In that purpose, the NAOS RTC computes over the image integration time the mean of the following data \citep{rabaud00}:
\begin{itemize}
\item at WFS frequency: Zernike Mode Mean, Zernike Mode Autocorrelation (128 first points, mean value not subtracted), Intensity Mean, Intensity Variance, Image Mean, Image Variance, Image pixel null/no-null count;
\item at 50 Hz: mean of the modal coefficients deduced from the residual slopes ($\bar{\epsilon}$, hereafter residual modal mean), covariance of the modal coefficients deduced from the residual slopes ($C_{\epsilon\epsilon}$, hereafter residual modal covariance matrix), mean of the modal coefficients deduced from the voltages and covariance of the modal coefficients deduced from the voltages (hereafter mirror modal mean and covariance matrix resp.). 
\end{itemize}
In practice, together with the CONICA FITS images, the observer gets:
\begin{itemize}
\item attached to the images: the two residual modal covariance and mirror modal covariance matrices and the corresponding two means;
\item written in the FITS header images: turbulence parameters, such as r$_0$, L$_0$ (cf. \citealt{fusco04} for details on their estimation), and AO loop parameters such as the Zernike mean noise $\overline{n^2_z}$.
\end{itemize}
\subsection{From the available data to the PSF}
\subsubsection{Computation of the mean residual phase OTF}
\label{sect:otfparal}
The WFS measurement,  $\hat{\epsilon}_\|$ can be decomposed into:
\vspace{-2mm}
\begin{equation}
\hat{\epsilon}_\|=\epsilon_\|+n+r
\label{eq20}
\end{equation}
\noindent where $\epsilon_\|$, $n$ and $r$ are the vectors representing the residual phase, the noise and the aliasing in the basis of the miror modes. Following \citet{veran97}, under the assumption of high temporal bandwidth, Eq.~\ref{eq20} leads to:
\vspace{-2mm}
\begin{equation}
\langle\epsilon_\|{\epsilon_\|}^t\rangle=\langle\hat{\epsilon}_\|{\hat{\epsilon}_\|}^t\rangle-\langle n n^t\rangle+\langle r r^t\rangle
\end{equation}

\noindent -- $\langle\hat{\epsilon}_\|{\hat{\epsilon}_\|}^t\rangle$ is directly obtained from the NAOS data: $\langle\hat{\epsilon}_\|{\hat{\epsilon}_\|}^t\rangle=C_{\epsilon\epsilon}-\langle\bar{\epsilon}{\bar{\epsilon}}^t\rangle$

\noindent -- $\langle r r^t\rangle$ is in a first step neglected. In the future, it will be computed by using a SH simulation of NAOS

\noindent -- $\langle n n^t\rangle$ was contemplated to be computed from the Zernike mean noise value $\overline{n^2_z}$ given in the CONICA image FITS header. To obtain this value, the vector of the residual phase is expressed in the Zernike mode basis, the open-loop measurement vector is reconstructed, the noise variances for each Zernike are calculated with the auto-correlation method  \citep{gendron95} and then summed and square-rooted. To take advantage of this Zernike mean noise value, one has to assume that the noise is not correlated from a subpupil to another and that the Zernike mean noise is equally distributed over all the considered Zernike modes (!). In that case, $\langle n n^t\rangle$ is given by (hereafter "$\overline{n^2_z}$ algorithm"):
\vspace{-2mm}
\begin{equation}
\langle n n^t\rangle=\frac{\displaystyle \overline{n^2_z}}{\displaystyle \sum_{i,j}M_{SZ_{i,j}}^2} M_{SM}M_{SM}^t 
\end{equation}

\noindent where $M_{SM}$ is the "slope to mirror mode" transformation matrix, $M_{SZ}$ the "slope to Zernike mode" transformation matrix and $M_{SZ\,i,j}$ its element along line $i$ and column $j$.

\subsubsection{Computation of the mean perpendicular phase OTF}
As proposed by \citet{veran97}, the mean perpendicular phase OTF can be computed from simulated phases screens, computed at $D/r_0=1$, of which one has subtracted their mirror components to obtain $\phi_{\epsilon_\perp}(\vec{x},t)$. These "perpendicular phase screens" have then to be scaled at the proper $D/r_0$ value by multiplying them by $(D/r_0)^{5/6}$. The $r_0$ value of the observations can be directly read in the CONICA image FITS header. Alternatively, it could be derived from the mirror modal mean and covariance matrix attached to the CONICA image FITS file. The former solution is presently adopted.

From these scaled perpendicular phase screens, one can derive the mean perpendicular phase OTF with two different algorithms:

\begin{itemize}
\item a "PSF-like" algorithm: one computes for each perpendicular phase screen the corresponding "instantaneous perpendicular PSF" by $\Big\|\mathcal{FFT}\big(\exp(i\phi_{\epsilon_\perp}(\vec{x},t)\big)\Big\|^2$, averages them to get $PSF_\perp(\vec{x},t)$ and, after a Fourier transform, the perpendicular OTF multiplied by the telescope OTF;
\item an "$U_{ij}$-like" algorithm: similarly to the commonly used computation of the $U_{ij}$ functions,
\begin{eqnarray}
\bar{D}_{\phi_{\epsilon_\perp}}(\vec{x},\vec{\rho})&=&\frac{\displaystyle\int P(\vec{x})P(\vec{x}+\vec{\rho})D_{\phi_{\epsilon_\perp}}(\vec{x},\vec{\rho})d\vec{x}}{\displaystyle\int P(\vec{x})P(\vec{x}+\vec{\rho})d\vec{x}}\\
&=&\frac{\displaystyle\left\langle \int P(\vec{x})P(\vec{x}+\vec{\rho})\left(\phi^2_{\epsilon_\perp}(\vec{x},t)+\phi^2_{\epsilon_\perp}(\vec{x}+\vec{\rho},t)-2 \phi_{\epsilon_\perp}(\vec{x},t) \phi_{\epsilon_\perp}(\vec{x}+\vec{\rho},t)\right)d\vec{x} \right\rangle }{\displaystyle\int P(\vec{x})P(\vec{x}+\vec{\rho})d\vec{x}}\\
&=&\frac{\displaystyle\left\langle Cor(\phi^2_{\epsilon_\perp} P,P)+Cor(P,\phi^2_{\epsilon_\perp} P)-2 Cor(\phi_{\epsilon_\perp} P,\phi_{\epsilon_\perp} P)\right\rangle}{\displaystyle Cor(P,P)}\\
&=&\frac{\displaystyle\mathcal{F}^{-1}\Big(\big\langle \mathcal{F}(\phi^2_{\epsilon_\perp} P)\mathcal{F}^*(P)+\mathcal{F}(P)\mathcal{F}^*(\phi^2_{\epsilon_\perp} P)-2\mathcal{F}(\phi_{\epsilon_\perp} P)\mathcal{F}^*(\phi_{\epsilon_\perp} P)\big\rangle\Big)}{\displaystyle\mathcal{F}^{-1}\left(\mid\mathcal{F}(P)\mid^2\right)}\\
&=&\frac{\displaystyle\mathcal{F}^{-1}\bigg(2 \Big\langle  \Re\left(\mathcal{F}(\phi^2_{\epsilon_\perp} P)\mathcal{F}^*(P)\right)-\mid\mathcal{F}(\phi_{\epsilon_\perp} P)\mid^2\Big\rangle \bigg)}{\displaystyle\mathcal{F}^{-1}\left(\mid\mathcal{F}(P)\mid^2\right)}
\end{eqnarray}
where ${\displaystyle Cor(f,g)=\int f(x).g(x+\rho) dx}$, $\mathcal{F}$ is the Fourier transform, $\mathcal{F}^{-1}$ the inverse Fourier transform, $\Re$ the real part of a complex value and $^*$ its conjugate. 
\end{itemize}
 \subsection{Contemplated NAOS software modifications}
\subsubsection{Description}
The Zernike mean noise value $\overline{n^2_z}$ is the only NAOS data available to compute the noise part of the OTF (Sect.~\ref{sect:naosdata}). Though, as explained in Sect.~\ref{sect:otfparal}, the assumptions made in this computation are fairly important and might lead to large uncertainties in the PSF estimation. It has lead us to propose to ESO two different RTC software modifications with different levels of impact on the software:

\begin{enumerate}
\item providing the vector of variance noises $n^2_{z_i}$ for all considered Zernikes, instead of their "mean" value $\overline{n^2_z}$ (hereafter "$n^2_{z_i}$ algorithm"). This is the modification that has the slightest impact on the software: since the vector of variance noise for all considered Zernikes is already computed in the software to derive $\overline{n^2_z}$, this software modification would only require to output the considered vector, in the CONICA image FITS header for example, instead of outputting $\overline{n^2_z}$. Always assuming that the noise is not correlated from a subpupil to another, we would first solve the equations $\displaystyle n^2_{s_i}=\sum_j M_{SZ_{i,j}}^2 n^2_{z_j}$ where $n^2_{s_i}$ is the i$^{th}$-element of the vector of variance noises for each slope mesurements. This can be done with a dedicated procedure, e.g. the IDL "constrained\_min.pro" procedure. $\langle n n^t\rangle$ is then derived using the  $M_{SM}$ "slope to mirror mode" transformation matrix,
\item providing the vector of variance noises $n^2_i$ for all considered mirror modes, i.e. the diagonal of $\langle n n^t\rangle$ (hereafter "$n^2_i$ algorithm"). In this larger software modification scheme, always assuming no correlation between the subpupils, we would first solve the equations $\displaystyle n^2_{s_i}=\sum_j M_{SM_{i,j}}^2 n^2_j$, similarly to what is proposed for the first software modification, and derive the entire $\langle n n^t\rangle$ covariance matrix with $M_{SM}$.
\end{enumerate}

\subsubsection{Test of the modifications}
\begin{figure}[t]
   \centering
   \includegraphics[width=5cm]{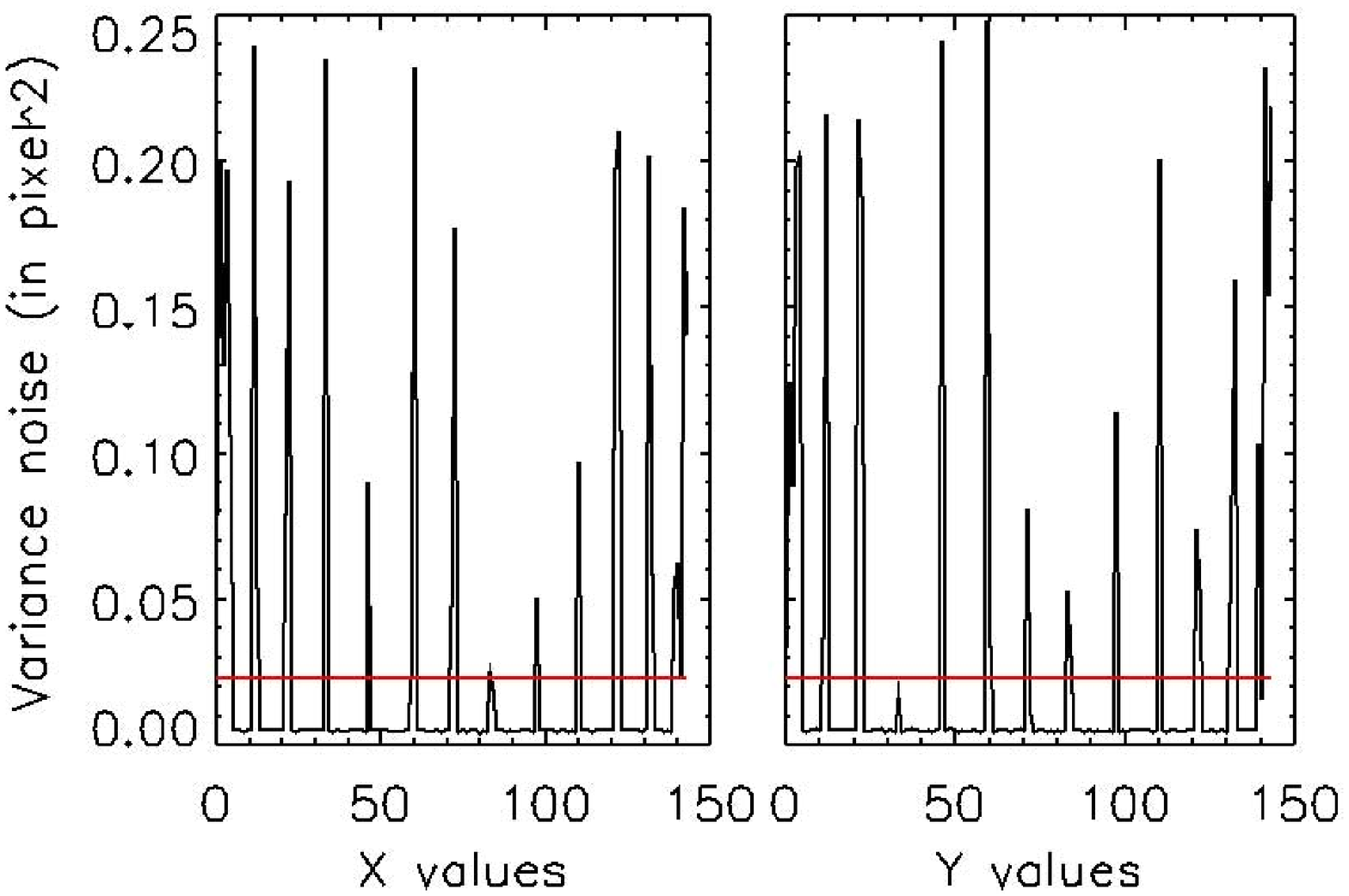}\hfill
   \includegraphics[width=5cm]{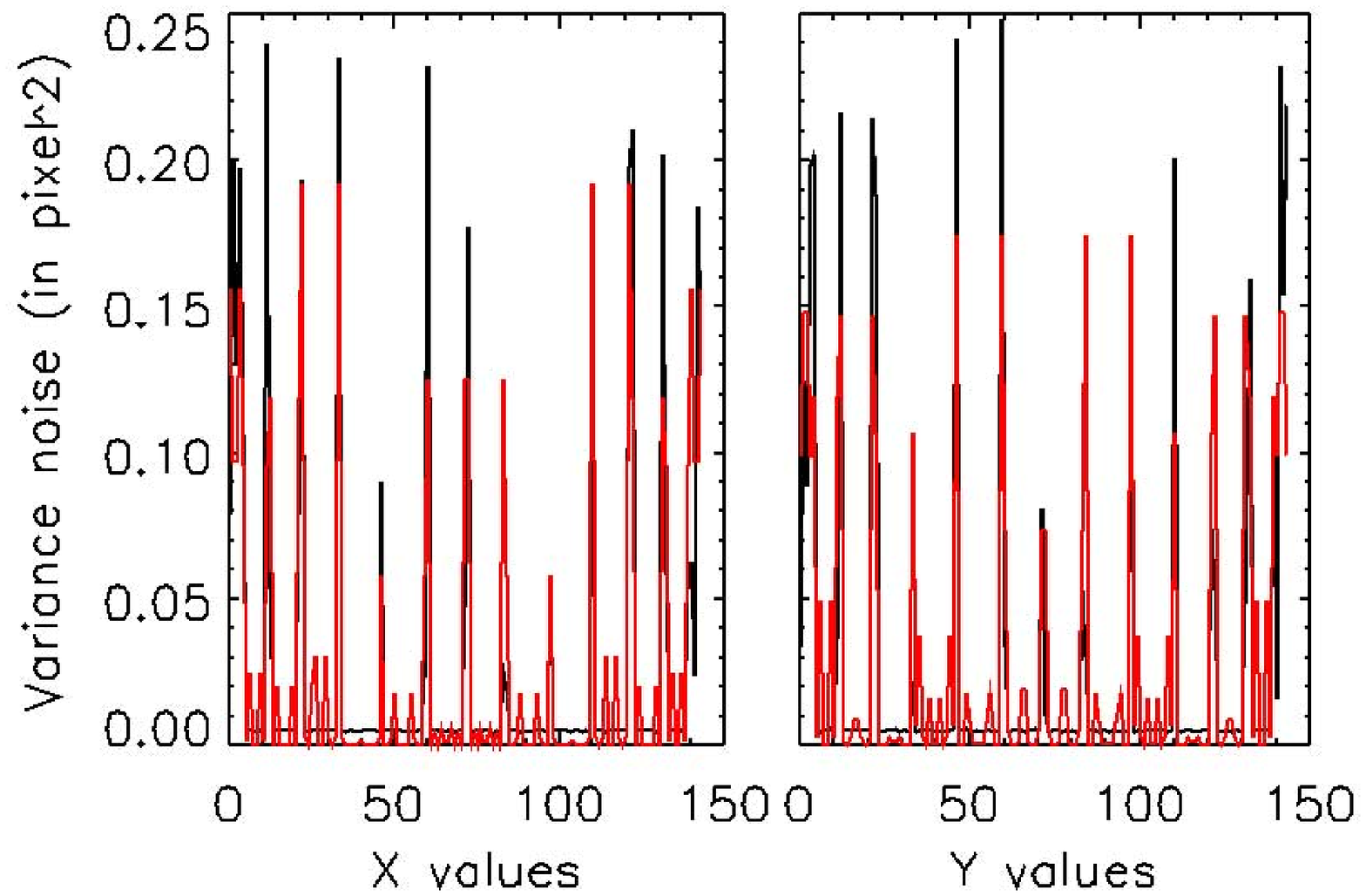}\hfill
   \includegraphics[width=5cm]{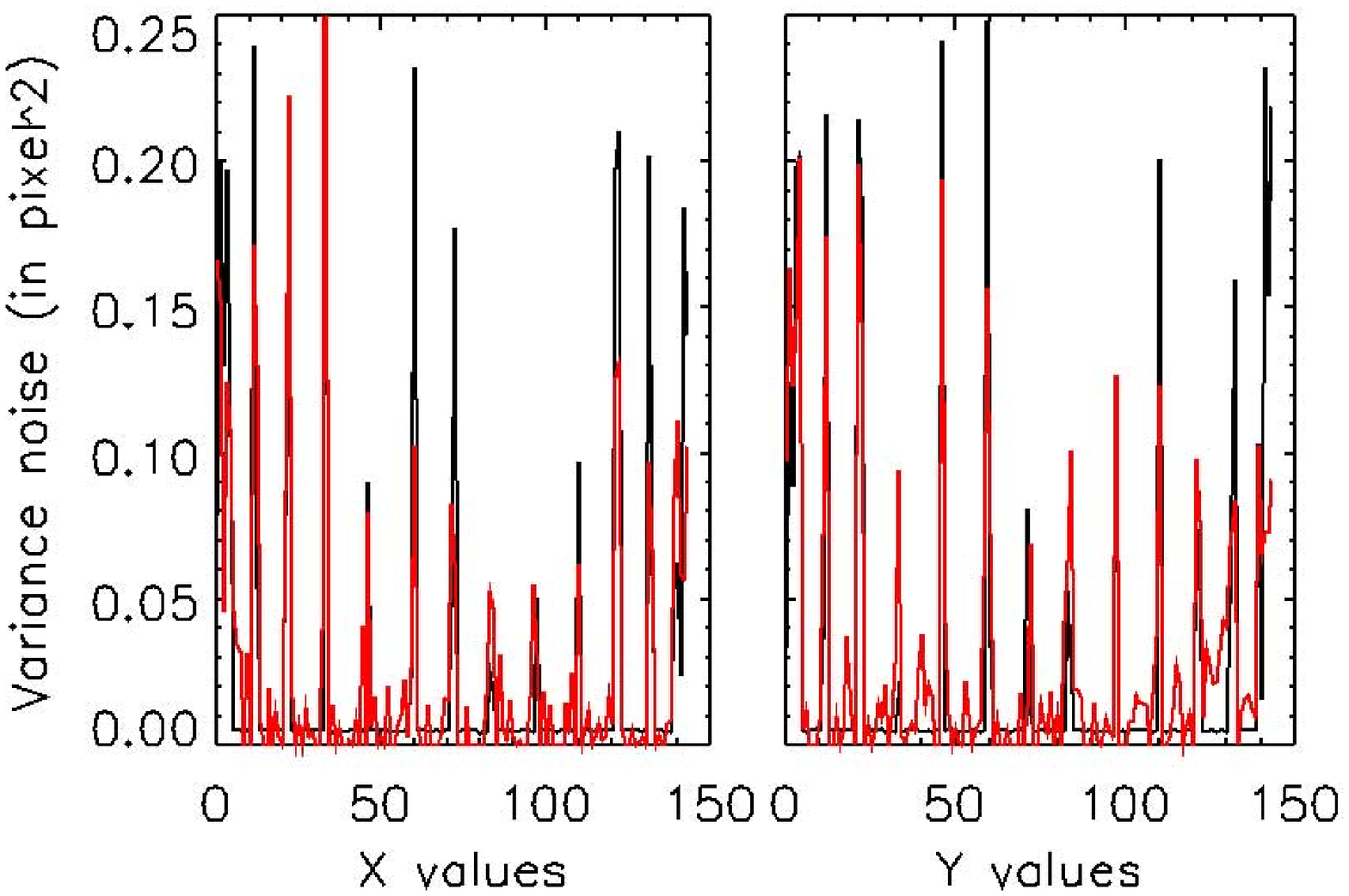}
      \caption{Drawn (black) and estimated (red) $n^2_{s_i}$ slope variance noises with the $\overline{n^2_z}$ (left), $n^2_{z_i}$ (middle) and $n^2_i$ (right) algorithms, for a given statistics distribution.}
         \label{fig:slopefit}
   \end{figure}

To test the contemplated modifications, we have drawn vectors of 144 $n^2_{s_i}$ slope variance noises with several statistics distribution, taking into account that the edge subpupils are more noisy than the central ones. Assuming no correlation between subpupils, i.e.  $\langle n_s n_s^t\rangle$ is diagonal and made of the $n^2_{s_i}$ elements, we have then:
\begin{itemize}
\item computed $\langle n n^t\rangle$ with the "slope to mode" transformation matrix, and then derived with the $V_{ii}$ algorithm the corresponding OTF (hereafter "reference OTF"), that we  estimate with the $\overline{n^2_z}$, $n^2_{z_i}$ and $n^2_i$ algorithms;
\item computed $\overline{n^2_z}$ with the "slope to Zernike" transformation matrix, used the "$\overline{n^2_z}$ algorithm" to get the estimated $n^2_{s_i}$ slope variance noises (Fig.~\ref{fig:slopefit} left), estimated $\langle n n^t\rangle$ with the "slope to mode" transformation matrix, and then derived with the $V_{ii}$ algorithm the corresponding estimated OTF;
\item computed $n^2_{z_i}$ with the "slope to Zernike" transformation matrix, used the "$n^2_{z_i}$ algorithm" to get the estimated $n^2_{s_i}$ slope variance noises (Fig.~\ref{fig:slopefit} middle),  estimated $\langle n n^t\rangle$ with the "slope to mode" transformation matrix, and then derived with the $V_{ii}$ algorithm the corresponding estimated OTF;
\item computed $n^2_i$ with the "slope to mode" transformation matrix, used the "$n^2_i$ algorithm" to get the estimated $n^2_{s_i}$ slope variance noises (Fig.~\ref{fig:slopefit} right), estimated $\langle n n^t\rangle$ with the "slope to mode" transformation matrix, and then derived with the $V_{ii}$ algorithm the corresponding estimated OTF;
\end{itemize}

\begin{figure}[!h]
   \centering
   \includegraphics[width=6cm]{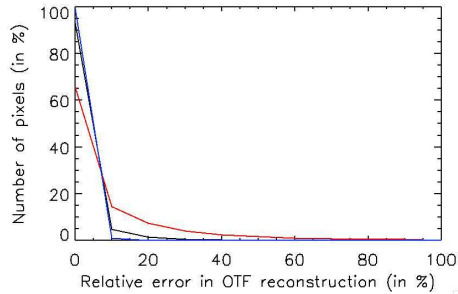}
      \caption{Histogram of the relative error between the "reference OTF" and the estimated OTF estimated with the $\overline{n^2_z}$ (upper red curve), $n^2_{z_i}$ (middle black curve) and $n^2_i$ (lower blue curve) algorithms, for the statistics distribution of the slope variance noises given in Fig.~\ref{fig:slopefit}.}
         \label{fig:histotf}
   \end{figure}

To measure the potential gain of the contemplated modifications, we have represented in Fig.~\ref{fig:histotf} the histogram, in number of pixels in the OTF, of the relative error between the "reference OTF" and the "estimated OTFs", for the statistics distribution of the slope variance noises shown in Fig.~\ref{fig:slopefit}. We thus show the substantial gain of the simplest software modification (providing the vector of variance noises $n^2_{z_i}$) compared to what is presently delivered by the NAOS RTC ($\overline{n^2_z}$). The gain of the second contemplated modification (providing the vector of variance noises $n^2_i$) seems to be poor with respect to the first one.

\section{Future work}
Regarding the new introduced algorithms, accurate AO simulations are needed to provide an estimation of $n$,  the equivalent number of independent realisations of PSFs whose sum has resulted in the observed PSF. The derived OTF variability would then be used in deconvolution algorithm, similarly to what has been done with PUEO and MISTRAL \citep{fusco99}. Concerning the NAOS PSF reconstruction software itself, the decision to proceed with the slightest software modification is still in discussion with ESO. Though, tests of the software with the present NAOS RTC data have just begun and more will be accumulated before a release to the community.  

\acknowledgments
This work has been initiated during my postdoc at ESO. It began with an attempt to adapt to NAOS the MPIA PSF reconstruction software: I would like to thank S. Egner and S. Hippler for this instructive and friendly collaboration. The large differences between ALFA and NAOS (modes, available wavefront-related measurements, etc) have lead to the elaboration of a NAOS-dedicated algorithm, within the ÒPHASEÓ collaboration, together with ESO and I am grateful to all those, at ONERA, ESO and Observatoire de Paris, who are involved in this work.

\end{document}